\documentclass[letterpaper, 10 pt, conference]{ieeeconf}  

\IEEEoverridecommandlockouts                              

\usepackage{amsmath} 
\usepackage{amssymb}  
\usepackage{graphicx}
\usepackage{subfig}
\newtheorem{remark}{Remark}

\title{\LARGE \bf
A Primal-Dual Method for Optimal Control and Trajectory Generation in High-Dimensional Systems*
}

\author{Matthew R. Kirchner$^{1,2}$, Gary Hewer$^{1}$, J\'{e}r\^{o}me Darbon$^{3}$, and Stanley Osher$^{4}$
\thanks{*This research was supported in part by the Office of Naval Research under Grant N00014-18-WX01382, Grant N00014-16-12119, and Grant N00014-16-12157, and in part by the Office of Naval Research under ILIR 4764 and ILIR 5100.}
\thanks{$^{1}$M. Kirchner and G. Hewer are with the Image and Signal Processing Branch, Research Directorate, Code 4F0000D, Naval Air Warfare Center Weapons Division, China Lake, CA 93555, USA  {\tt\small \{matthew.kirchner, gary.hewer\}@navy.mil}}%
\thanks{$^{2}$M. Kirchner is with the Electrical and Computer Engineering Department, University of California, Santa Barbara, CA 93106, USA {\tt\small kirchner@ucsb.edu}}
\thanks{$^{3}$J. Darbon is with the Division of Applied Mathematics, Brown University,         Providence, RI 02912, USA         {\tt\small jerome\_darbon@brown.edu}}
\thanks{$^{4}$S. Osher is with the Department of Mathematics, University of California,         Los Angeles, CA 90095, USA         {\tt\small sjo@math.ucla.edu}}}

\begin{document}

\maketitle
\thispagestyle{empty}
\pagestyle{empty}

\begin{abstract}
Presented is a method for efficient computation of the Hamilton\textendash Jacobi
(HJ) equation for time-optimal control problems using the generalized
Hopf formula. Typically, numerical methods to solve the HJ equation
rely on a discrete grid of the solution space and exhibit exponential
scaling with dimension. The generalized Hopf formula avoids the use
of grids and numerical gradients by formulating an unconstrained convex
optimization problem. The solution at each point is completely independent,
and allows a massively parallel implementation if solutions at multiple
points are desired. This work presents a primal-dual method for efficient
numeric solution and presents how the resulting optimal trajectory
can be generated directly from the solution of the Hopf formula, without
further optimization. Examples presented have execution times on the
order of milliseconds and experiments show computation scales approximately
polynomial in dimension with very small high-order coefficients. 
\end{abstract}

\section{Introduction}

Hamilton\textendash Jacobi equations play a fundamental role in optimal
control theory as they establish sufficient conditions for optimality
\cite{osmolovskii1998calculus}. Traditionally, numerical solutions
to HJ equations require a dense, discrete grid of the solution space
\cite{osher2006level,mitchell2008flexible,mitchell2005time}. Computing
the elements of this grid scales poorly with dimension and has limited
use for problems with dimension greater than four. The exponential
dimensional scaling in optimization is sometimes referred to as the
``curse of dimensionality'' \cite{bellman2015adaptive,bellman1957dynamic}.
Recent research \cite{darbon2016algorithms,kirchner2017time} has
discovered numerical solutions based on the generalized Hopf formula
that do not require a grid and can be used to efficiently compute
solutions of a certain class of Hamilton\textendash Jacobi PDEs that
arise in linear control theory and differential games. 

A key hurdle in the development of efficient high-dimensional solutions
is the time-dependent Hamiltonian that results for general control
problems. Kirchner et al. \cite{kirchner2017time} applied the generalized
Hopf formula with time-dependent Hamiltonian to efficiently solve
multi-vehicle optimal pursuit-evasion using the linearized models
found in \cite{palumbo2010modern}. In that work, the unique, specific
structure of the model was used to derive a closed form solution to
the gradient of the objection function, thereby allowing efficient
optimization. This same technique cannot be applied to general linear
systems and attempts to use numeric gradient approximation would increase
computation time.

Darbon and Osher presented a proximal splitting algorithm in \cite{darbon2016algorithms}
using the split\textbf{ }Bregman/ADMM approach \cite{goldstein2009split},
but this only applies to systems with time-independent Hamiltonians
of the form $\dot{x}=f\left(u\left(t\right)\right)$, and has limited
use for general linear control problems. Chow et al. \cite{chow2016timealgorithm}
developed a coordinate descent method, but this optimization method
lacks robustness for the nonsmooth optimization that typically result
from the Hopf formula for optimal control problems.

This work presents a parallel proximal splitting optimization method
\cite{combettes2011proximal} for solving time-optimal control problems
with the generalized Hopf formula, including those with time-dependent
Hamiltonians. This allows efficient solutions to generalized linear
models, even when no explicit gradient of the objective function is
known and without resorting to consensus-type algorithms \cite{parikh2014proximal}.
Section \ref{sec:Solutions-to-the} reviews using the Hopf formula
for solutions to the Hamilton\textendash Jacobi equations that arise
in optimal linear control and largely follows the work of \cite{kirchner2017time}
and \cite{darbon2016algorithms}. The main contributions of this paper
are presented in Section \ref{sec:Proximal-Splitting-Methods}, with
a primal-dual method for solving the Hopf formula, Section \ref{sec:Time-Optimal-Control},
which presents obtaining the optimal control, and Section \ref{sec:Trajectory-Generation-with},
where the optimal trajectory can be obtained directly from the solution
of the Hopf formula. The new methods are applied on various time-optimal
control problems and are presented in Section \ref{sec:Results}. 

\section{\label{sec:Solutions-to-the}Solutions to Hamilton\textendash Jacobi
Equations with the Hopf Formula}

Consider system dynamics represented as
\begin{equation}
\dot{x}\left(t\right)=f\left(u\left(t\right)\right),\label{eq: Basic diff eq}
\end{equation}
where $x\left(t\right)\in\mathbb{R}^{n}$ is the system state and
$u\left(t\right)\in\mathcal{U}\subset\mathbb{R}^{m}$ is the control
input, constrained to lie in the convex admissible control set $\mathcal{U}$.
The system in $\left(\ref{eq: Basic diff eq}\right)$ describes how
the state evolves in time and is considered a dynamic constraint when
control inputs $u\left(t\right)$ are to be optimized. We consider
a cost functional for a given initial time $t$, and terminal time
$T$ 
\begin{equation}
K\left(x,u,t\right)=\int_{t}^{T}L\left(u\left(s\right)\right)ds+J\left(x\left(T\right)\right),\label{eq: Cost Function}
\end{equation}
where $x\left(T\right)$ is the solution of $\left(\ref{eq: Basic diff eq}\right)$
at terminal time, $T$. We assume that the terminal cost function
$J:\mathbb{R}^{n}\rightarrow\mathbb{R}$ is convex. The function $L:\mathbb{R}^{m}\rightarrow\mathbb{R}\cup\left\{ +\infty\right\} $
is the running cost, and represents the rate that cost is accrued.
The value function $v:\mathbb{R}^{n}\times(-\infty,T]\rightarrow\mathbb{R}$
is defined as the minimum cost, $K$, among all admissible controls
for a given state $x$, and time $t\leq T$ with
\begin{equation}
v\left(x,t\right)=\underset{u\in\mathcal{U}}{\text{inf}}\,K\left(x,u,t\right).\label{eq: Value function}
\end{equation}
The value function in $\left(\ref{eq: Value function}\right)$ satisfies
the dynamic programming principle \cite{bryson1975applied,evans10}
and also satisfies the following initial value Hamilton\textendash Jacobi
(HJ) equation by defining the function $\varphi:\mathbb{R}^{n}\times\mathbb{R}\rightarrow\mathbb{R}$
as $\varphi\left(x,t\right)=v\left(x,T-t\right)$, with $\varphi$
being the viscosity solution of
\begin{equation}
\begin{cases}
\frac{\partial\varphi}{\partial t}\left(x,t\right)+H\left(\nabla_{x}\varphi\left(x,t\right)\right)=0 & \text{in}\,\mathbb{R}^{n}\times\left(0,+\infty\right),\\
\varphi\left(x,0\right)=J\left(x\right) & \forall x\in\mathbb{R}^{n},
\end{cases}\label{eq:Initial value HJ PDE}
\end{equation}
where the Hamiltonian $H:\mathbb{R}^{n}\rightarrow\mathbb{R}\cup\left\{ +\infty\right\} $
is defined by
\begin{equation}
H\left(p\right)=\underset{c\in\mathbb{R}^{m}}{\text{sup}}\left\{ \left\langle -f\left(c\right),p\right\rangle -L\left(c\right)\right\} .\label{eq: Basic Hamiltonian definition}
\end{equation}
We proceed with the Hamilton Jacobi formulation for time-optimal control
to reach some convex terminal set $\Omega$, though the following
methods can be generalized to other optimization problems. To apply
the constraint that the control must be bounded, we introduce the
following running cost $L=\mathcal{I}_{\mathcal{U}}$, where $\mathcal{I}_{\mathcal{C}}:\mathbb{R}^{n}\rightarrow\mathbb{R}\cup\left\{ +\infty\right\} $
is the indicator function for the set $\mathcal{C}$ and is defined
by
\[
\mathcal{I}_{\mathcal{C}}\left(x\right)=\begin{cases}
0 & \text{if}\,x\in\mathcal{C}\\
+\infty & \text{otherwise.}
\end{cases}
\]
 This reduces the Hamiltonian to 
\[
H\left(p\right)=\underset{c\in\mathcal{U}}{\text{max}}\left\langle -f\left(c\right),p\right\rangle .
\]
Solving the HJ equation $\left(\ref{eq:Initial value HJ PDE}\right)$
describes how the value function evolves with time at any point in
the state space, and from this optimal control policies can be found.

\subsection{Viscosity Solutions with the Hopf Formula}

It was shown in \cite{evans10} that an exact, point-wise viscosity
solution to $\left(\ref{eq:Initial value HJ PDE}\right)$ can be found
using the Hopf formula \cite{hopf1965generalized}. Moreover, no discrete
grid is constructed, and the formula can provide a numerical method
that is efficient even when the state space is high-dimensional. The
value function can be found with the Hopf formula
\begin{equation}
\varphi\left(x,t\right)=-\underset{p\in\mathbb{R}^{n}}{\text{min}}\left\{ J^{\star}\left(p\right)+tH\left(p\right)-\left\langle x,p\right\rangle \right\} ,\label{eq: Basic Hopf formula}
\end{equation}
where the Fenchel\textendash Legendre transform $g^{\star}:\mathbb{R}^{n}\rightarrow\mathbb{R}\cup\left\{ +\infty\right\} $
of a convex, proper, lower semicontinuous function $g:\mathbb{R}^{n}\rightarrow\mathbb{R}\cup\left\{ +\infty\right\} $
is defined by \cite{hiriart2012fundamentals} 
\begin{equation}
g^{\star}\left(p\right)=\underset{x\in\mathbb{R}^{n}}{\text{sup}}\left\{ \left\langle p,x\right\rangle -g\left(x\right)\right\} .\label{eq: Fenchel transform}
\end{equation}

\subsection{General Linear Models\label{subsec:General-Linear-Models}}

Now consider the following linear state space model
\begin{equation}
\dot{x}\left(t\right)=Ax\left(t\right)+Bu\left(t\right),\label{eq:general linear system}
\end{equation}
with $A\in\mathbb{R}^{n\times n}$, $B\in\mathbb{R}^{n\times m}$,
state vector $x\in\mathbb{R}^{n}$, and control input $u\in\mathbb{\mathcal{U}\subset R}^{m}$.
We can make a change of variables
\begin{equation}
z\left(t\right)=e^{-tA}x\left(t\right),\label{eq:change of varibles}
\end{equation}
which results in the following system
\begin{equation}
\dot{z}\left(t\right)=e^{-tA}Bu\left(t\right),\label{eq:z transformed system}
\end{equation}
with terminal cost function now defined in $z$ with
\begin{equation}
\varphi\left(z,0\right)=J_{z}\left(z,0\right)=J_{x}\left(e^{TA}z\right).\label{eq:Terminal cost as function of z}
\end{equation}
For clarity in the sections to follow, we use the notation $H_{z}$
to refer to the Hamiltonian for systems defined by $\left(\ref{eq:z transformed system}\right)$,
and $H_{x}$ for systems defined by $\left(\ref{eq:general linear system}\right)$.
Additionally, with a slight abuse of notation, we denote by $J_{x}^{\star}\left(p,T\right)$
the Fenchel transform of $J\left(x,t\right)$ with respect to the
variable $x$ at time $t=T$. Notice that the system $\left(\ref{eq:z transformed system}\right)$
is now time-varying, and it was shown in \cite[Section 5.3.2, p. 215]{kurzhanski2014dynamics}
that the Hopf formula can be generalized for a time-dependent Hamiltonian
to solve for the value function of the system in $\left(\ref{eq:z transformed system}\right)$
with
\begin{align}
\varphi\left(z_{0},T\right) & =-\underset{p\in\mathbb{R}^{n}}{\text{min}}\Bigg\{ J_{z}^{\star}\left(p,0\right)\label{eq:generalized hopf formula}\\
 & +\int_{0}^{T}H_{z}\left(p,s\right)ds-\left\langle z_{0},p\right\rangle \Bigg\},\nonumber 
\end{align}
with $H_{z}$ defined as
\begin{equation}
H_{z}\left(p,t\right)=\underset{c\in\mathcal{U}}{\text{max}}\left\langle -e^{-\left(T-t\right)A}Bc,p\right\rangle .\label{eq:Hamiltonian max form, linear}
\end{equation}
The change of variable to $\left(T-t\right)$ is required for time
since the problem was converted to an initial value formulation from
a terminal value formulation in $\left(\ref{eq:Initial value HJ PDE}\right)$.
The value function found by solving the unconstrained optimization
problem in $\left(\ref{eq:generalized hopf formula}\right)$ can be
thought of as the minimal cost of a system starting at initial state
$z\left(0\right)=z_{0}$ and ending at terminal state $z\left(T\right)$. 
\begin{remark}
While we are solving an initial value problem in this work, we can
solve for a candidate solution of a two point boundary value problem
(TPBVP) by selecting for the terminal set $\Omega$ a ball, and shrinking
the radius until we get arbitrarily close to the terminal boundary
condition of the corresponding TPBVP \cite[Section 2.7.2, p. 66]{Mitchell_toolbox}.
\end{remark}

\section{\label{sec:Proximal-Splitting-Methods}Proximal Splitting Methods
for Control}

Proximal splitting methods \cite{combettes2011proximal} are a powerful
group of convex optimization algorithms that efficiently solve non-smooth
minimization problems in high dimensions. These methods are used for
problems of the form
\[
\underset{p\in\mathbb{R}^{n}}{\min}\,G\left(p\right)+F\left(p\right),
\]
where proximal points for $G$ and $F$ can be easily computed. These
family of methods have been proven effective for image processing
and compressed sensing applications \cite{goldstein2009split}. The
proximal point of $f$ at $p$ for some $\alpha>0$ is given by
\begin{align*}
\left(I+\alpha\partial f\right)^{-1}\left(p\right) & =\text{arg\,}\underset{w\in\mathbb{R}^{n}}{\text{min}}\left\{ \alpha f\left(w\right)+\frac{1}{2}\left\Vert w-p\right\Vert _{2}^{2}\right\} .
\end{align*}
The primal-dual algorithm \cite{chambolle2011first} is a proximal
splitting algorithm that solves the minimization problem of the form
\begin{equation}
\underset{p\in\mathbb{R}^{n}}{\min}\,G\left(p\right)+F\left(Kp\right),\label{eq: primal CP}
\end{equation}
with $K\in\mathbb{R}^{m\times n}$ and $G,F$ being assumed convex,
by converting $\left(\ref{eq: primal CP}\right)$ to the saddle-point
problem
\[
\underset{p\in\mathbb{R}^{n}}{\min}\,\underset{y\in\mathbb{R}^{m}}{\max}\left\langle Kp,y\right\rangle +G\left(p\right)-F^{\star}\left(y\right).
\]
The minimizer can be found by iterating the following update procedure
\begin{equation}
\begin{cases}
y^{k+1} & =\left(I+\sigma\partial F^{\star}\right)^{-1}\left(y^{k}+\sigma K\bar{p}^{k}\right)\\
p^{k+1} & =\left(I+\tau\partial G\right)^{-1}\left(p^{k}-\tau K^{\top}y^{k+1}\right)\\
\bar{p}^{k+1} & =p^{k+1}+\theta\left(p^{k+1}-p^{k}\right),
\end{cases}\label{eq: PD iterates}
\end{equation}
until convergence with $\tau,\sigma>0$ being the primal and dual
step sizes and $\theta\in\left[0,1\right]$. It was shown in \cite{chambolle2011first}
that the primal-dual method of $\left(\ref{eq: PD iterates}\right)$
converges at a rate of $\mathcal{O}\left(1/k\right)$ for general
convex functions $F$ and $G$ provided the condition $\tau\sigma\left\Vert K\right\Vert ^{2}<1$
is satisfied. If one or both of $F$ and $G$ are strongly convex,
then \cite{chambolle2011first} provides an alteration to the algorithm
in $\left(\ref{eq: PD iterates}\right)$ that was shown to converge
at a faster rate. While some problems in Section \ref{sec:Results}
meet this criteria, these accelerated algorithms are outside the scope
of this work, and all examples used the algorithm in $\left(\ref{eq: PD iterates}\right)$.

\subsection{Primal-Dual Solutions to the Generalized Hopf Formula}

Suppose $\mathcal{U}$ is a closed convex set such that $0\in\text{int}\,\mathcal{U}$, where $\text{int}\,\mathcal{U}$ denotes the interior of the set $\mathcal{U}$.
Then $\left(\mathcal{I}_{\mathcal{U}}\right)^{\star}$ defines a norm
$\left\Vert \left(\cdot\right)\right\Vert $ and we denote by $\left\Vert \left(\cdot\right)\right\Vert _{*}$
its dual norm \cite{hiriart2012fundamentals}. Also consider that
the set $\mathcal{U}$ can be scaled by an injective linear transformation,
$Q^{-1}$, to give the appropriate problem-specific control bound,
then $\left(\ref{eq:Hamiltonian max form, linear}\right)$ can be
written as
\begin{equation}
H_{z}\left(p,t\right)=\left\Vert \left(-e^{-\left(T-t\right)A}BQ\right)^{\top}p\right\Vert _{*}.\label{eq:time-varying Hamiltonian}
\end{equation}
For simplicity, we follow \cite{kirchner2017time} by approximating
the integral in $\left(\ref{eq:generalized hopf formula}\right)$
with a left Riemann sum quadrature with $N$ equally spaced terms
defined by
\begin{equation}
t_{i}=i\Delta t,\label{eq:Left riemann time samples}
\end{equation}
with $i\in\left\{ 0,\ldots,N-1\right\} $ and $\Delta t=\frac{T}{N}$.
The generalized Hopf formula, for some terminal time $T$, becomes
\begin{align}
\varphi\left(z_{0},T\right) & =-\underset{p\in\mathbb{R}^{n}}{\text{min}}\Bigg\{ J_{z}^{\star}\left(p,0\right)\nonumber \\
 & +\Delta t\sum_{i=0}^{N-1}\left\Vert \left(-e^{-\left(T-t_{i}\right)A}BQ\right)^{\top}p\right\Vert _{*}\nonumber \\
 & -\left\langle z_{0},p\right\rangle \Bigg\}.\label{eq: Generalized Hopf with quadrature}
\end{align}
Other forms of quadrature can be used for integral approximations,
such as trapezoidal or Simpson's rule \cite{mathews2004numerical},
that increase approximation accuracy without sacrificing computational
performance. Additionally, other advanced approximations could be
considered, such as those employed by pseudospectral methods in \cite{ross2012review},
and will be investigated in future work.

To formulate as a primal-dual optimization, first set
\begin{equation}
G\left(p\right)=J_{z}^{\star}\left(p,0\right)-\left\langle z_{0},p\right\rangle .\label{eq:G, for prox}
\end{equation}
Now define 
\begin{equation}
K_{i}=\left(-e^{-\left(T-t_{i}\right)A}BQ\right)^{\top}\label{eq:K_i}
\end{equation}
 and $F_{i}=\Delta t\left\Vert \left(\cdot\right)\right\Vert _{*}=\left(\mathcal{I}_{\Delta t\mathcal{U}}\right)^{\star}$,
where we denote by $\mathcal{I}_{\alpha\mathcal{C}}$ the indicator
function of a set $\mathcal{C}$ scaled by a constant $\alpha>0$
defined as
\[
\mathcal{I}_{\mathcal{\alpha C}}\left(x\right)=\begin{cases}
0 & \text{if}\,\frac{x}{\alpha}\in\mathcal{C}\\
+\infty & \text{otherwise.}
\end{cases}
\]
 The sum term in $\left(\ref{eq: Generalized Hopf with quadrature}\right)$
can be written as
\[
\Delta t\sum_{i=0}^{N-1}\left\Vert \left(-e^{-\left(T-t_{i}\right)A}BQ\right)^{\top}p\right\Vert _{*}=\sum_{i=0}^{N-1}F_{i}\left(K_{i}p\right).
\]
We can now form the matrix $K$ as
\[
K=\left[\begin{array}{c}
K_{1}\\
K_{2}\\
\vdots\\
K_{N}
\end{array}\right],
\]
which gives
\begin{equation}
F\left(Kp\right)=\sum_{i=0}^{N-1}F_{i}\left(K_{i}p\right),\label{eq: F as sum}
\end{equation}
and combined with $\left(\ref{eq:G, for prox}\right)$ is now in the
form of $\left(\ref{eq: primal CP}\right)$. Note that $K$ in this
formulation is non square (preconditioning \cite{pock2011diagonal}
can be used to enhance convergence rate). If the action of the matrix
exponential in $\left(\ref{eq:K_i}\right)$ is not known, then it
can be quickly evaluated, for all time samples, without resorting
to computing the matrix exponential with \cite{al2011computing}.
The structure of $\left(\ref{eq: F as sum}\right)$ forms a \emph{separable
sum}. This implies that for some $y=\left(y_{1},y_{2},\ldots,y_{N}\right)$
\[
F^{\star}\left(y_{1},y_{2},\ldots,y_{N}\right)=F_{1}^{\star}\left(y_{1}\right)+F_{2}^{\star}\left(y_{2}\right)+\cdots+F_{N}^{\star}\left(y_{N}\right).
\]
Recall that $F_{i}=\left(\mathcal{I}_{\Delta t\mathcal{U}}\right)^{\star}$,
thus
\begin{equation}
F_{i}^{\star}\left(y_{i}\right)=\mathcal{I}_{\Delta t\mathcal{U}}.\label{eq:F_i}
\end{equation}
The proximal operator of a separable sum is simply
\begin{equation}
\left(I+\partial F^{\star}\right)^{-1}\left(y_{1},y_{2},\ldots,y_{N}\right)=\left[\begin{array}{c}
\left(I+\partial F_{1}^{\star}\right)^{-1}\left(y_{1}\right)\\
\left(I+\partial F_{2}^{\star}\right)^{-1}\left(y_{2}\right)\\
\vdots\\
\left(I+\partial F_{N}^{\star}\right)^{-1}\left(y_{N}\right)
\end{array}\right].\label{eq:seperable prox}
\end{equation}
The proximal operators in $\left(\ref{eq:seperable prox}\right)$
are independent of each other, and as a result can be computed in
parallel. This can be advantageous for real-time implementations in
hardware such as multi-core embedded CPUs and field programmable gate
arrays (FPGAs).

\subsection{Stopping Criteria}

Care must be taken to select the stopping criterion of the algorithm
in $\left(\ref{eq: PD iterates}\right)$. The step sizes $\tau$ and
$\sigma$ are in general not equal and any stopping criteria must
account for this. As a result we choose the primal and dual residuals
\cite[See Eqs. 10 and 12]{goldstein2013adaptive} as a step-size dependent
stopping criteria and stop iterating $\left(\ref{eq: PD iterates}\right)$
when the conditions
\[
\left\Vert \left(p^{k}-p^{k-1}\right)/\tau-K^{\top}\left(y^{k}-y^{k-1}\right)\right\Vert <\varepsilon,
\]
and
\[
\left\Vert \left(y^{k}-y^{k-1}\right)/\sigma-K\left(p^{k}-p^{k-1}\right)\right\Vert <\varepsilon,
\]
are both met. For all the experiments
listed in Section \ref{sec:Results}, $\epsilon$ is set to $10^{-4}$.

\section{Time-Optimal Control\label{sec:Time-Optimal-Control}}

To find the time-optimal control to some convex terminal set $\Omega$,
choose a convex terminal cost function $J_{x}$ such that
\[
\begin{cases}
J_{x}\left(x\right)<0 & \text{for any}\,x\in\text{int}\,\Omega,\\
J_{x}\left(x\right)>0 & \text{for any}\,x\in\left(\mathbb{R}^{n}\setminus\Omega\right),\\
J_{x}\left(x\right)=0 & \text{for any}\,x\in\left(\Omega\setminus\text{int}\,\Omega\right),
\end{cases}
\]
where $\text{int}\,\Omega$ denotes the interior of $\Omega$. The
intuition behind defining the terminal cost function this way is simple.
If the value function $\varphi\left(x_{0},T\right)<0$ for some $x_{0}$
and $T$, then there exists a control $u\left(t\right)$ that drives
the state from the initial condition at $x_{0}$ to the final state,
$x\left(T\right)$, inside the set $\Omega$. The smallest value of
time $T$, such that $\varphi\left(x_{0},T\right)=0$ is the minimum
time to reach the set $\Omega$, starting at state $x_{0}$. Recall
that the initial value function, $J_{z}\left(p,0\right)$, and its
associated Fenchel\textendash Legendre transform, $J_{z}^{\star}\left(p,0\right)$
of the Hopf formula in $\left(\ref{eq:generalized hopf formula}\right)$,
is defined in $z$, and must be transformed with $\left(\ref{eq:Terminal cost as function of z}\right)$.
The minimum time to reach $\Omega$ is denoted by $T^{*},$ and the
control computed at $T^{*}$ is the time-optimal control. As first
noted in \cite{lions1986hopf}, Hopf formula is itself a Fenchel\textendash Legendre
transform. It follows from a well known property of the Fenchel\textendash Legendre
transform \cite{darbon2015convex} that the unique minimizer of $\left(\ref{eq:generalized hopf formula}\right)$
is the gradient of the value function
\begin{align}
\nabla_{z}\varphi\left(z_{0},T\right) & =\text{arg}\,\underset{p\in\mathbb{R}^{n}}{\text{min}}\Bigg\{ J_{z}^{\star}\left(p,0\right)\label{eq:gradient as argmin}\\
 & +\int_{0}^{T}H_{z}\left(p,s\right)ds-\left\langle z_{0},p\right\rangle \Bigg\},\nonumber 
\end{align}
provided the gradient exists. So when solving for the value function
using $\left(\ref{eq:generalized hopf formula}\right)$, we automatically
solve for the gradient. We will refer to the minimizer in $\left(\ref{eq:gradient as argmin}\right)$
as $p^{*}=\nabla_{z}\left(z_{0},T\right)$.

We propose solving for the minimum time to reach the set $\Omega$, $T^{*}$, by a hybrid
method of the bisection method and Newton's method. Newton's method
has been shown to have faster convergence (quadratic) than bisection,
but is unstable when the gradient is small and motivates the use of
a hybrid method. We can iterate time, $t_{n+1}$, with Newton's as
\begin{equation}
t_{n+1}=t_{n}-\frac{\varphi\left(x_{0},t_{n}\right)}{\frac{\partial\varphi}{\partial t}\left(x_{0},t_{n}\right)}.\label{eq:basic newton update}
\end{equation}
As noted in \cite{darbon2016algorithms}, $\frac{\partial\varphi}{\partial t}\left(x_{0},t\right)$
must satisfy the Hamilton\textendash Jacobi equation $\left(\ref{eq:Initial value HJ PDE}\right)$.
Therefore we have
\[
\frac{\partial\varphi}{\partial t}\left(x_{0},t_{n}\right)=-H_{x}\left(\nabla_{x}\varphi\left(x_{0},t_{n}\right),x_{0}\right).
\]
We also see from $\left(\ref{eq:change of varibles}\right)$ and applying
the chain rule that
\begin{align*}
\nabla_{x}\varphi\left(z\left(t\right),T-t\right) & =\nabla_{x}\varphi\left(e^{-tA}x\left(t\right),T-t\right)\\
 & =e^{-tA^{\top}}\nabla_{z}\left(z\left(t\right),T-t\right).
\end{align*}
Therefore when $t=0$, then $z_{0}=x_{0}$, $\varphi\left(z_{0},T\right)=\varphi\left(x_{0},T\right)$,
and
\[
\nabla_{x}\left(x_{0},T\right)=\nabla_{z}\left(z_{0},T\right).
\]
This implies that $\left(\ref{eq:basic newton update}\right)$ can
be written as
\begin{equation}
t_{n+1}=t_{n}+\frac{\varphi\left(z_{0},t_{n}\right)}{H_{x}\left(\nabla_{z}\varphi\left(z_{0},t_{n}\right),z_{0}\right)}.\label{eq:basic newton update with hamiltonian}
\end{equation}
For the purpose of evaluating $\left(\ref{eq:basic newton update with hamiltonian}\right)$,
there is no need to apply the change of variables as in $\left(\ref{eq:change of varibles}\right)$.
Therefore we have
\[
H_{x}\left(p^{*},z_{0}\right)=-z_{0}^{\top}A^{\top}p^{*}+\left\Vert -Q^{\top}B^{\top}p^{*}\right\Vert _{*}.
\]
If it is known that a single zero exists on the interval $T^{*}\in\left[0,t_{max}\right]$
then we use the Newton update from $\left(\ref{eq:basic newton update with hamiltonian}\right)$
to find $t_{n+1}$. With the value function computed at each Newton
iteration, we can keep track of the updated interval $T^{*}\in\left[t_{min},t_{max}\right]$,
and use a bisection update if the Newton update of $t_{n+1}$ is out
side this interval. Once the minimum time to reach the set $\Omega$, $T^{*}$, is found,
the optimal control $u^{*}\left(t\right)$ can be found from the relation
\begin{align}
\nabla_{p}H_{z}\left(\nabla_{z}\varphi\left(z_{0},T^{*}\right),T^{*}\right) & =\nabla_{p}H_{z}\left(p^{*},T^{*}\right)\nonumber \\
 & =e^{-\left(T^{*}-t\right)A}Bu^{*}\left(t\right).\label{eq:Optimal control}
\end{align}

\section{\label{sec:Trajectory-Generation-with}Trajectory Generation with
the Generalized Hopf Formula}

Using the solution of the Hopf formula, we illustrate a dynamic programming
point of view \cite{bellman1957dynamic,evans10} of the associated
Hamilton\textendash Jacobi equation to compute the optimal trajectory.
We denote by $\gamma_{z}\left(s\right)\in\mathbb{R}^{n}$, with $s\in\left(0,T\right)$,
as the state trajectory with $\gamma_{z}\left(0\right)=z_{0}$. Recall
the fact that the solution of short-time Hopf formula $\left(\ref{eq: Basic Hopf formula}\right)$
is itself a Fenchel transform \cite{lions1986hopf} 
\begin{equation}
\varphi\left(z,t\right)=\left(J^{\star}+tH\right)^{\star}\left(z\right).\label{eq:Hopf is Fenchel}
\end{equation}
Note that the minimizer of $J_{z}^{\star}\left(p\right)$ is the optimal
terminal state $\gamma_{z}\left(T\right)$, and, if $J_{z}^{\star}$
is differentiable, then can be found with
\begin{equation}
\gamma_{z}\left(T\right)=\frac{\partial}{\partial p}J_{z}^{\star}\left(p^{*},0\right),\label{eq:terminal points}
\end{equation}
where $p^{*}$ is the solution to the Hopf formula in $\left(\ref{eq: Basic Hopf formula}\right)$.
Now consider the case where $p^{*}$ is the solution to the generalized
Hopf formula given in $\left(\ref{eq:generalized hopf formula}\right)$,
and what follows is an analysis in the variable $z$. This implies
that
\begin{align}
\varphi\left(z_{0},T\right) & =-\Bigg\{ J_{z}^{\star}\left(p^{*},0\right)+\Delta t\sum_{i=0}^{N-1}H_{z}\left(p^{*},t_{i}\right)\label{eq:value of generalized hopf formula}\\
 & -\left\langle z_{0},p^{*}\right\rangle \Bigg\},\nonumber 
\end{align}
where each of the quadrature time samples $t_{i}$ are equally spaced
by $\Delta t$ seconds on the interval $\left[0,T\right]$ as defined
by $\left(\ref{eq:Left riemann time samples}\right)$. Recall from
Section \ref{subsec:General-Linear-Models} that we denote by $\varphi_{z}^{\star}\left(p,T\right)$
the Fenchel transform of $\varphi\left(z,t\right)$ with respect to
the variable $z$ at time $t=T$. If we write the Hopf formula with
initial convex data $J\left(z,T-\Delta t\right)=\varphi\left(z,T-\Delta t\right)$,
then we can find the level set evolution only for a short time, $\Delta t$,
starting at value $\varphi\left(z,T-\Delta t\right)$ with 
\begin{align*}
\varphi\left(z_{0},T\right) & =-\Bigg\{\varphi_{z}^{\star}\left(p^{*},T-\Delta t\right)\\
 & =+\Delta tH_{z}\left(p^{*},T-\Delta t\right)-\left\langle z_{0},p^{*}\right\rangle \Bigg\}.
\end{align*}
Following $\left(\ref{eq:terminal points}\right)$, the optimal state,
with respect to $z$ is
\begin{equation}
\gamma_{z}\left(\Delta t\right)=\frac{\partial}{\partial p}\varphi_{z}^{\star}\left(p^{*},T-\Delta t\right).\label{eq: next to last optimal trajectory}
\end{equation}
From $\left(\ref{eq:Hopf is Fenchel}\right)$ we conclude that
\begin{align}
\varphi_{z}^{\star}\left(p^{*},T-\Delta t\right) & =\left(J^{\star}+\Delta tH_{z}\right)_{p}^{\star\star}\left(z,T-\Delta t\right)\nonumber \\
 & =J_{z}^{\star}\left(p^{*},T-2\Delta t\right)\label{eq:recursion}\\
 & +\Delta tH_{z}\left(p^{*},T-2\Delta t\right).\nonumber 
\end{align}
The last line in $\left(\ref{eq:recursion}\right)$ is due to the
fact that if $f$ is convex, proper and lower semicontinuous, then
$f^{\star\star}=f$. We form a recursive operation, repeating $\left(\ref{eq:recursion}\right)$
until we reach time zero and get
\begin{align*}
\varphi\left(z_{0},T\right) & =-\Bigg\{\varphi_{z}^{\star}\left(p^{*},0\right)\\
 & +\Delta t\sum_{i=0}^{N-1}H_{z}\left(p^{*},t_{i}\right)-\left\langle z_{0},p^{*}\right\rangle \Bigg\},
\end{align*}
which is equivalent to $\left(\ref{eq:value of generalized hopf formula}\right)$
since by definition $J\left(x,0\right)=\varphi\left(x,0\right)$.
This suggests that the generalized Hopf formula with the integral
in $\left(\ref{eq:generalized hopf formula}\right)$ approximated
by quadrature is equivalent to the composition of many short-time
Hopf formulas $\left(\ref{eq: Basic Hopf formula}\right)$ of length
$\Delta t$. Also, the recursion can be applied to the optimal terminal
point from $\left(\ref{eq: next to last optimal trajectory}\right)$
as
\begin{align*}
\gamma_{z}\left(\Delta t\right) & =\frac{\partial}{\partial p}\varphi_{z}^{\star}\left(p^{*},T-\Delta t\right)\\
 & =\frac{\partial}{\partial p}\left\{ J_{z}^{\star}\left(p^{*},0\right)+\Delta t\sum_{i=0}^{N-1}H_{z}\left(p^{*},t_{i}\right)\right\} .
\end{align*}
This can equivalently used for the optimal trajectory at any time
sample $t_{i}$ as
\begin{align}
\gamma_{z}\left(t_{i}\right) & =\frac{\partial}{\partial p}\left\{ J_{z}^{\star}\left(p^{*},0\right)+\Delta t\sum_{k=0}^{i}H_{z}\left(p^{*},t_{k}\right)\right\} \label{eq:optimal trajectory at sample}\\
 & =\frac{\partial}{\partial p}J_{z}^{\star}\left(p^{*},0\right)+\Delta t\sum_{k=0}^{i}\frac{\partial}{\partial p}H_{z}\left(p^{*},t_{k}\right).\nonumber 
\end{align}
If we are interested in the trajectory at each quadrature time sample,
we don't have to recompute the sum for each $t_{k}$, since we can
incrementally build the trajectory point from $t_{k+1}$. Note that
each point in $\left(\ref{eq:optimal trajectory at sample}\right)$
is in terms of the state variable $z$, and can be found for $x$
by applying inverse of the transform given in $\left(\ref{eq:change of varibles}\right)$.
Non-rigorously we see that the time rate of change of the state trajectory
in $\left(\ref{eq:optimal trajectory at sample}\right)$ is equal
to the gradient with respect to $p$ of the Hamiltonian, which satisfies
Pontryagin's Maximum Principal \cite{ross2015primer}, though it was
derived using only the Hopf formula and basic principals of convex
analysis.

\section{\label{sec:Results}Results}

\begin{figure}
\begin{centering}
\includegraphics[height=6cm]{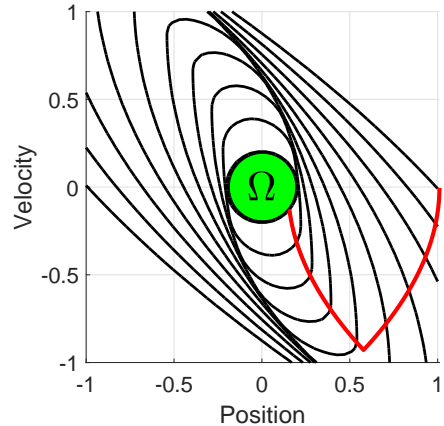}
\par\end{centering}
\caption{The double integrator example. The zero level set evolution solved
at various times shown in black. Direct trajectory generation shown
in red. The terminal set, $\Omega$, is shaded green. \label{fig:The-level-set}}
\end{figure}

The primal-dual method presented in Section \ref{sec:Proximal-Splitting-Methods}
was implemented in MATLAB R2017a on a laptop equipped with an Intel
Core i7-7500 CPU running at 2.70 GHz. For all experiments, $100$
time samples were used for the quadrature in $\left(\ref{eq: Generalized Hopf with quadrature}\right)$
and the dual step size was set to $\sigma=\frac{1}{\tau\left\Vert K\right\Vert ^{2}}$,
and $\theta=1$. The initial conditions for all examples is $p^{0}=\bar{p}^{0}=x_{0}$
for the primal variable and $y^{0}=Kp^{0}$ for the dual variable.
Note that for some of the examples to follow, the action of the matrix
exponential is known in closed form and that can be used for increased
computational enhancement. Since in general this is not the case,
we used \cite{al2011computing} to numerically compute the action
of the matrix exponential as to show dimensional scaling properties
even for the most general case. 

\subsection{Double Integrator}

We begin with the simple double integrator problem with system $\dot{x}=Ax+Bu$
with
\[
A=\left[\begin{array}{cc}
0 & 1\\
0 & 0
\end{array}\right]
\]
and
\[
B=\left[\begin{array}{c}
0\\
1
\end{array}\right],
\]
where the state $x\in\mathbb{R}^{2}$ is position and velocity. This
problem is selected since closed form optimal solutions exist and
is low enough dimension to compare the level set evolution to that
of grid based numeric techniques such as that in \cite{mitchell2008flexible,mitchell2005toolbox}.
We consider the control to be constrained to $u\in\left[-1,1\right]$
and implies a control of $\mathcal{C}=\left\{ u:\left|u\right|\leq1\right\} $.
After a change of variables $\left(\ref{eq:change of varibles}\right)$,
the Hamiltonian becomes
\[
H_{z}\left(p,t\right)=\left|\left(-e^{-\left(T-t\right)A}B\right)^{\top}p\right|.
\]
We chose the terminal set to be an ellipsoidal with
\begin{equation}
\Omega=\left\{ x:\left\langle x,W^{-1}x\right\rangle \leq1\right\} .\label{eq:Capture set}
\end{equation}
where $W$ is symmetric positive definite. For the initial cost function
$J_{x}$, the elements of $W$ are selected such that $\Omega$ is
a circle with radius $r=0.2$. The terminal cost function becomes
\[
J_{z}\left(z,0\right)=\left\langle z,V\left(0\right)z\right\rangle -1
\]
where $V\left(t\right)=e^{\left(T-t\right)A^{\top}}W^{-1}e^{\left(T-t\right)A}$.
This gives
\begin{equation}
J_{z}^{\star}\left(p,0\right)=\frac{1}{4}\left\langle p,V\left(0\right)^{-1}p\right\rangle +1.\label{eq: terminal cost in z}
\end{equation}
In this example $\left(\ref{eq:G, for prox}\right)$ becomes
\begin{equation}
G\left(p\right)=\frac{1}{4}\left\langle p,V\left(0\right)^{-1}p\right\rangle -\left\langle z_{0},p\right\rangle +1,\label{eq:G split for example}
\end{equation}
which is quadratic and results in the following proximal point of
$G$ at $p$:
\begin{equation}
\left(I+\tau\partial G\right)^{-1}\left(p\right)=\left(I+\frac{1}{2}\tau V\left(0\right)^{-1}\right)^{-1}\left(p+\tau z_{0}\right).\label{eq:prox G for example}
\end{equation}
Note that we do not need to compute the inverse of $V\left(0\right)$
since
\[
V\left(0\right)^{-1}=e^{-TA}We^{-TA^{\top}}.
\]
Likewise, with $F_{i}^{\star}$ defined in $\left(\ref{eq:F_i}\right)$,
the proximal points of each $F_{i}^{\star}$ at $y_{i}$ in $\left(\ref{eq:seperable prox}\right)$
is given by
\[
\left(I+\sigma\partial F_{i}^{\star}\right)^{-1}\left(y_{i}\right)=\Delta t\,\text{sign}\left(w_{i}\right)\text{min}\left(\left|w_{i}\right|,1\right),
\]
where $w_{i}=\frac{y_{i}}{\Delta t}$ and $\text{sign}\left(\beta\right)=1$
if $\beta\geq0$ and $-1$ otherwise.

We computed the solution to the Hamilton\textendash Jacobi equation
at each point on a grid, $\left[-1,1\right]^{2}$, of 50 equally spaced
points in each dimension. This was done to average execution time
for a large number of initial conditions. The average computational
time was $2.1\,ms$ per point on the grid and the zero contours of
the value function for ten different times equally spaced on $t\in[0,T^{*}]$
are shown in Figure \ref{fig:The-level-set}. The value of $T^{*}$
we set as by solving for the minimum time to the zero level set for
the initial state $z_{0}=x_{0}=\left(1,0\right)^{\top}$ using the
method presented in Section \ref{sec:Time-Optimal-Control}. The primal
step size was set to $\tau=10$. 

The optimal trajectory starting at $z_{0}$ was computed following
$\left(\ref{eq:optimal trajectory at sample}\right)$ with 
\[
\gamma_{z}\left(T^{*}\right)=\frac{\partial}{\partial p}J_{z}^{\star}\left(p^{*},0\right)=\frac{1}{2}V\left(0\right)^{-1}p^{*},
\]
since $J_{x}\left(x\right)$ is quadratic. The gradient of the Hamiltonian
is 
\begin{align*}
\frac{\partial}{\partial p}H_{z}\left(p^{*},t_{k}\right) & =-e^{-\left(T^{*}-t_{k}\right)A}B\\
 & \times\text{sign}\left(\left(-e^{-\left(T^{*}-t_{k}\right)A}B\right)^{\top}p^{*}\right).
\end{align*}
The trajectory as computed in $\left(\ref{eq:optimal trajectory at sample}\right)$
is shown in red in Figure \ref{fig:The-level-set}.

\subsection{Unscented Optimal Control\label{subsec:Unscented-Optimal-Control}}

\begin{figure*}
\begin{centering}
\subfloat[100 random samples of initial conditions with the existing control
formulation. 66 samples reached the goal state and 34 missed.\label{fig:Without UT plot}]{\includegraphics[height=5cm]{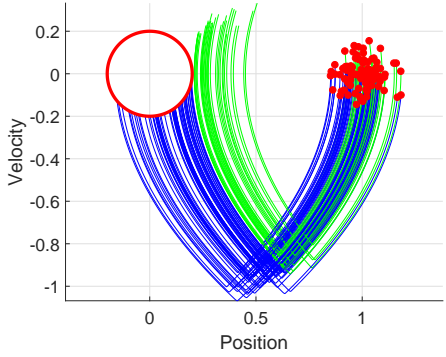}

}~~\subfloat[100 random samples of initial conditions with uncertainty modeled
as an unscented control problem. 79 samples reached the goal state
and 21 missed.\label{fig:With UT plot}]{\begin{centering}
\includegraphics[height=5cm]{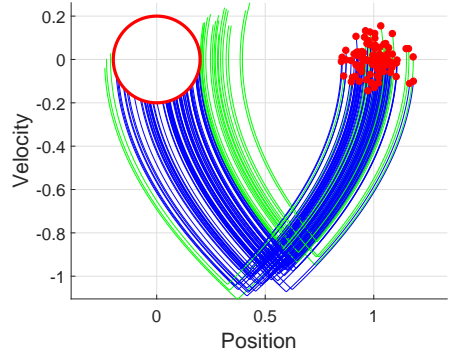}
\par\end{centering}

}
\par\end{centering}
\caption{An example of unscented control applied to a time-optimal double integrator
problem with uncertain initial state. Red are samples of the initial
states. Blue trajectories are random initial states that reached the
goal state and green are trajectories that missed the goal state.
Best viewed in color. \label{fig:UT Example}}
\end{figure*}

We can utilize the favorable dimensional scaling of the proposed methods
to generate solutions that are robust to system uncertainty. These
uncertainties could be the initial state or system parameters such
as mass, lift coefficients, or other properties. The idea is to augment
the system to include samples that represent different initial conditions
or parameters that share one, common control input. This system forms
a tychastic \cite{aubin2012stochastic} differential equation, since
the parameters are fixed, but unknown at run time. The goal is to
select a single control so the aggregate cost of all samples is optimized. 

For uncertainties that can be modeled as Gaussian, we can choose the
samples deterministically with the unscented transform \cite{julier2002scaled},
and these samples are typically referred to in literature as sigma
points. This transform provides a second-order approximation of the
moments of a Gaussian distribution propagated through a nonlinear
function. It was developed for state estimation problems and in this
context is known as the unscented Kalman filter \cite{julier2004unscented,thrun2005probabilistic}.
Using the unscented transform for sample selection to approximate
the tychastic optimal control problem was developed by Ross et al.
\cite{ross2015unscented,ross2016path}. The extra state dimensions
that result from this technique is not as problematic with proximal
splitting as it may be with other methods. 

Let $x=\left(\chi^{\left[1\right]},\chi^{\left[2\right]},\ldots,\chi^{\left[2n+1\right]}\right)^{\top}$
, with $\chi^{\left[i\right]}\in\mathbb{R}^{n}$ represent the new
state vector augmented by $2n+1$ states generated from the unscented
transform and the new system becomes
\[
\dot{x}=\left[\begin{array}{c}
\dot{\chi}_{1}\\
\dot{\chi}_{2}\\
\vdots\\
\dot{\chi}_{2n+1}
\end{array}\right]=\left[\begin{array}{c}
f_{1}\left(\chi^{\left[1\right]},u\right)\\
f_{2}\left(\chi^{\left[2\right]},u\right)\\
\vdots\\
f_{2n+1}\left(\chi^{\left[2n+1\right]},u\right)
\end{array}\right],
\]
subject to initial condition $z_{0}=x_{0}=\left(\chi_{0}^{\left[1\right]},\chi_{0}^{\left[2\right]},\ldots,\chi_{0}^{\left[2n+1\right]}\right)$.
Take for example the problem of uncertainty in initial condition,
$x_{0}\sim\mathcal{N}\left(\mu,\Sigma\right)$. As an unscented control
problem, the dynamics are the same, so $f_{1}=f_{2}=\cdots=f_{2n+1}=f$,
and our augmented system is $\dot{x}=\tilde{A}x+\tilde{B}u$ with
\[
\tilde{A}=\left[\begin{array}{cccc}
A & \cdots &  & 0\\
\vdots & A &  & \vdots\\
 &  & \ddots\\
0 & \cdots &  & A
\end{array}\right],
\]
and
\[
\tilde{B}=\left[\begin{array}{c}
B\\
B\\
\vdots\\
B
\end{array}\right].
\]
The initial state becomes $x_{0}=\left\{ \chi_{0}^{\left[i\right]}\right\} _{i=1,\ldots,2n+1}$,
where each $\chi_{0}^{\left[i\right]}$ is formed by the unscented
transform with mean $\mu$ and covariance $\Sigma$. The mean square
error of the terminal state relative to some goal state, $\hat{x}_{T}$
is found with the unscented transform by
\begin{equation}
\text{MSE}\approx\sum_{i=1}^{2n+1}w_{m}^{\left[i\right]}\left(\chi_{T}^{\left[i\right]}-\hat{x}_{T}\right)^{\top}\left(\chi_{T}^{\left[i\right]}-\hat{x}_{T}\right),\label{eq:UT recondtructed Cov}
\end{equation}
where $w_{m}$ is the mean weight factor\footnote{For more information on the generation of the unscented sigma points,
and their weights, see \cite{thrun2005probabilistic}.}, and $\chi_{T}^{\left[i\right]}$ is the terminal state for the $i$-th
unscented sigma point. For this example, we wish to find the minimum
time to reach the origin subject to the constraint that the mean square
error is less than some threshold $\ell$. This can be found from
$\left(\ref{eq:UT recondtructed Cov}\right)$ with
\[
\text{MSE}=\mathbb{E}\left[\left(\hat{x}_{T}-x\left(T\right)\right)^{\top}\left(\hat{x}_{T}-x\left(T\right)\right)\right]\leq\ell.
\]
We select the origin as the goal with $\hat{x}=\left(0,0\right)^{\top}$,
and formulate as an unscented control problem. The trace of the terminal
covariance can be represented by the quadratic
\begin{equation}
J\left(z\right)=\left\langle z,V\left(0\right)z\right\rangle -\ell\label{eq:Trace quadratic cost}
\end{equation}
 with $V\left(t\right)=e^{\left(T-t\right)A^{\top}}W^{-1}e^{\left(T-t\right)A}$
and 
\[
W=\left[\begin{array}{ccc}
\left(w_{m}^{\left[1\right]}\right)^{-1}I_{2} & \cdots & 0\\
\vdots & \ddots & \vdots\\
0 & \cdots & \left(w_{m}^{\left[2n+1\right]}\right)^{-1}I_{2}
\end{array}\right],
\]
where $I_{2}$ is the $2\times2$ identity matrix.

Figure \ref{fig:UT Example} shows example trajectories when the initial
condition has random perturbations, with $\mu=x_{0}$ and $\Sigma=\pi^{2}I_{2}$,
for a double integrator problem. The standard deviation was set to
$\pi=0.0667$. If the initial state is \emph{exactly} what was used
to compute the optimal control, then the trajectory generated reaches
the goal and is time-optimal. However, if the initial state is perturbed,
some trajectories miss the intended goal entirely. In this particular
example, it is especially sensitive to perturbations to the ``right''
in the spatial $x$ direction and is shown in the Figure \ref{fig:Without UT plot}
on the left. Of $100$ random initial states, $34$ miss the goal
state. For the resulting $10$ dimensional unscented control problem
with $\tau=0.5$, the HJ solutions were found on average $4.0\,ms$
per point. Trajectory samples are shown in the Figure \ref{fig:Without UT plot}
on the right, the number of trajectories that miss the goal is reduced
to only $5$. 

\subsection{\label{subsec:Dimensional-Scaling}Dimensional Scaling}

\begin{figure}
\centering{}\includegraphics[width=8cm]{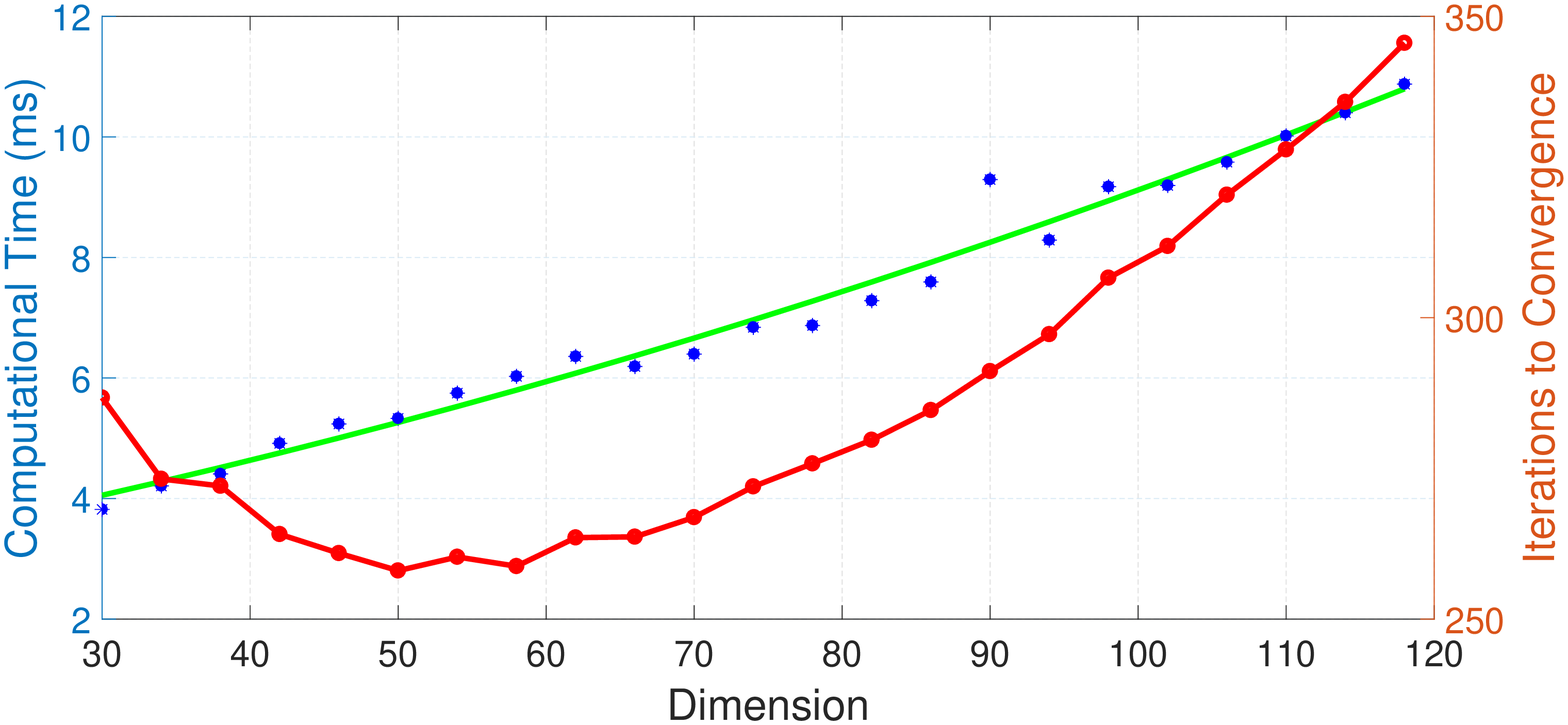}\caption{Dimensional scaling properties as described by the experiment in Section
\ref{subsec:Dimensional-Scaling}. The blue dots represent average
computational time for an experiment and the green line is the least
squares polynomial fit to computational time. The red displays average
iterations to convergence. \label{fig:Dimensional-scaling-properties}}
\end{figure}

Next we seek to analyze how the proposed method scales with dimension.
We can construct a problem similar to that presented in Section \ref{subsec:Unscented-Optimal-Control}
but by selecting samples at random as opposed to using the unscented
transform. Constructing a problem in this fashion is something not
typically done in practice, but allows us to vary the number of random
samples, and hence alter the dimension of the problem in a consistent
and uniform way. The initial state for $k$ samples becomes $x_{0}=\left\{ \chi_{0}^{\left[i\right]}\right\} _{i=1,\ldots,k}$,
where each $\chi_{0}^{\left[i\right]}$ is an independent and identically
distributed (iid) random vector drawn according to $\chi_{0}^{\left[i\right]}\sim\mathcal{N}\left(\mu,\Sigma\right)$.
To again penalize the trace, the terminal cost function is the same
as $\left(\ref{eq:Trace quadratic cost}\right)$ except $W$ is now
defined by
\[
W=\left[\begin{array}{ccc}
kI_{2} & \cdots & 0\\
\vdots & \ddots & \vdots\\
0 & \cdots & kI_{2}
\end{array}\right].
\]
Figure \ref{fig:Dimensional-scaling-properties} shows the average
computational time and average iterations to convergence for dimensions
ranging from $30$ to $120$. The green line in the figure is the
least squares polynomial fit for average computation time in milliseconds.
If we let $d$ denote problem dimension for the experiments, the fit
is $t_{\text{comp}}=2.382\times10^{-4}d^{2}+0.0414d+2.598$. Note
the extremely small quadratic coefficient. 

\section{Conclusion }

Presented is a parallel primal-dual method to solve the Hamilton\textendash Jacobi
equation for time-optimal control using the generalized Hopf formula.
We empirically showed how the method scales approximately quadratic
with dimension, though with small quadratic coefficient. The experiments
were shown using Matlab, and simple implementation in a compiled
language could provide significant computational improvement. Future
work includes increased experimentation with different systems, the
use of more advanced quadrature methods, adaptive step sizes, time-varying
systems, and state dependent Hamiltonians.

\bibliographystyle{IEEEtran}
\bibliography{WP}

\begin{thebibliography}{10}
\providecommand{\url}[1]{#1}
\csname url@samestyle\endcsname
\providecommand{\newblock}{\relax}
\providecommand{\bibinfo}[2]{#2}
\providecommand{\BIBentrySTDinterwordspacing}{\spaceskip=0pt\relax}
\providecommand{\BIBentryALTinterwordstretchfactor}{4}
\providecommand{\BIBentryALTinterwordspacing}{\spaceskip=\fontdimen2\font plus
\BIBentryALTinterwordstretchfactor\fontdimen3\font minus
  \fontdimen4\font\relax}
\providecommand{\BIBforeignlanguage}[2]{{%
\expandafter\ifx\csname l@#1\endcsname\relax
\typeout{** WARNING: IEEEtran.bst: No hyphenation pattern has been}%
\typeout{** loaded for the language `#1'. Using the pattern for}%
\typeout{** the default language instead.}%
\else
\language=\csname l@#1\endcsname
\fi
#2}}
\providecommand{\BIBdecl}{\relax}
\BIBdecl

\bibitem{osmolovskii1998calculus}
N.~P. Osmolovskii, \emph{Calculus of Variations and Optimal Control}.\hskip 1em
  plus 0.5em minus 0.4em\relax American Mathematical Society, 1998, vol. 180.

\bibitem{osher2006level}
S.~Osher and R.~Fedkiw, \emph{Level Set Methods and Dynamic Implicit
  Surfaces}.\hskip 1em plus 0.5em minus 0.4em\relax Springer Science \&
  Business Media, 2006, vol. 153.

\bibitem{mitchell2008flexible}
I.~M. Mitchell, ``The flexible, extensible and efficient toolbox of level set
  methods,'' \emph{Journal of Scientific Computing}, vol.~35, no.~2, pp.
  300--329, 2008.

\bibitem{mitchell2005time}
I.~Mitchell, A.~M. Bayen, and C.~J. Tomlin, ``A time-dependent
  {Hamilton}-{Jacobi} formulation of reachable sets for continuous dynamic
  games,'' \emph{IEEE Transactions on Automatic Control}, vol.~50, no.~7, pp.
  947--957, 2005.

\bibitem{bellman2015adaptive}
R.~E. Bellman, \emph{Adaptive Control Processes: A Guided Tour}.\hskip 1em plus
  0.5em minus 0.4em\relax Princeton University Press, 2015.

\bibitem{bellman1957dynamic}
------, \emph{Dynamic Programming}.\hskip 1em plus 0.5em minus 0.4em\relax
  Princeton University Press, 1957, vol.~1, no.~2.

\bibitem{darbon2016algorithms}
J.~Darbon and S.~Osher, ``Algorithms for overcoming the curse of dimensionality
  for certain {Hamilton}-{Jacobi} equations arising in control theory and
  elsewhere,'' \emph{Research in the Mathematical Sciences}, vol.~3, no.~1,
  p.~19, 2016.

\bibitem{kirchner2017time}
M.~R. Kirchner, R.~Mar, G.~Hewer, J.~Darbon, S.~Osher, and Y.~T. Chow,
  ``Time-optimal collaborative guidance using the generalized {Hopf} formula,''
  \emph{IEEE Control Systems Letters}, vol.~2, no.~2, pp. 201--206, 2018.

\bibitem{palumbo2010modern}
N.~F. Palumbo, R.~A. Blauwkamp, and J.~M. Lloyd, ``Modern homing missile
  guidance theory and techniques,'' \emph{Johns Hopkins APL Technical Digest},
  vol.~29, no.~1, pp. 42--59, 2010.

\bibitem{goldstein2009split}
T.~Goldstein and S.~Osher, ``The split {Bregman} method for {L1}-regularized
  problems,'' \emph{SIAM Journal on Imaging Sciences}, vol.~2, no.~2, pp.
  323--343, 2009.

\bibitem{chow2016timealgorithm}
Y.~T. Chow, J.~Darbon, S.~Osher, and W.~Yin, ``Algorithm for overcoming the
  curse of dimensionality for time-dependent non-convex {Hamilton--Jacobi}
  equations arising from optimal control and differential games problems,''
  \emph{Journal of Scientific Computing}, pp. 1--27, 2016.

\bibitem{combettes2011proximal}
P.~L. Combettes and J.-C. Pesquet, ``Proximal splitting methods in signal
  processing,'' in \emph{Fixed-point Algorithms for Inverse Problems in Science
  and Engineering}.\hskip 1em plus 0.5em minus 0.4em\relax Springer, 2011, pp.
  185--212.

\bibitem{parikh2014proximal}
N.~Parikh and S.~Boyd, ``Proximal algorithms,'' \emph{Foundations and Trends in
  Optimization}, vol.~1, no.~3, pp. 127--239, 2014.

\bibitem{bryson1975applied}
A.~R. Bryson and Y.-C. Ho, \emph{Applied Optimal Control: Optimization,
  Estimation and Control}.\hskip 1em plus 0.5em minus 0.4em\relax CRC Press,
  1975.

\bibitem{evans10}
L.~C. Evans, \emph{Partial differential equations}.\hskip 1em plus 0.5em minus
  0.4em\relax Providence, R.I.: American Mathematical Society, 2010.

\bibitem{hopf1965generalized}
E.~Hopf, ``Generalized solutions of non-linear equations of first order,''
  \emph{Journal of Mathematics and Mechanics}, vol.~14, pp. 951--973, 1965.

\bibitem{hiriart2012fundamentals}
J.-B. Hiriart-Urruty and C.~Lemar{\'e}chal, \emph{Fundamentals of convex
  analysis}.\hskip 1em plus 0.5em minus 0.4em\relax Springer Science \&
  Business Media, 2012.

\bibitem{kurzhanski2014dynamics}
A.~B. Kurzhanski and P.~Varaiya, \emph{Dynamics and Control of Trajectory
  Tubes: Theory and Computation}.\hskip 1em plus 0.5em minus 0.4em\relax
  Springer, 2014, vol.~85.

\bibitem{Mitchell_toolbox}
I.~M. Mitchell, ``A toolbox of level set methods,'' UBC Department of Computer
  Science, Tech. Rep. TR-2007-11, 2007.

\bibitem{chambolle2011first}
A.~Chambolle and T.~Pock, ``A first-order primal-dual algorithm for convex
  problems with applications to imaging,'' \emph{Journal of Mathematical
  Imaging and Vision}, vol.~40, no.~1, pp. 120--145, 2011.

\bibitem{mathews2004numerical}
J.~H. Mathews and K.~D. Fink, \emph{Numerical Methods using {MATLAB}}.\hskip
  1em plus 0.5em minus 0.4em\relax Prentice Hall, 1999, vol.~3.

\bibitem{ross2012review}
I.~M. Ross and M.~Karpenko, ``A review of pseudospectral optimal control: From
  theory to flight,'' \emph{Annual Reviews in Control}, vol.~36, no.~2, pp.
  182--197, 2012.

\bibitem{pock2011diagonal}
T.~Pock and A.~Chambolle, ``Diagonal preconditioning for first order
  primal-dual algorithms in convex optimization,'' in \emph{2011 IEEE
  International Conference on Computer Vision (ICCV)}.\hskip 1em plus 0.5em
  minus 0.4em\relax IEEE, 2011, pp. 1762--1769.

\bibitem{al2011computing}
A.~H. Al-Mohy and N.~J. Higham, ``Computing the action of the matrix
  exponential, with an application to exponential integrators,'' \emph{SIAM
  Journal on Scientific Computing}, vol.~33, no.~2, pp. 488--511, 2011.

\bibitem{goldstein2013adaptive}
T.~Goldstein, M.~Li, X.~Yuan, E.~Esser, and R.~Baraniuk, ``Adaptive primal-dual
  hybrid gradient methods for saddle-point problems,'' \emph{arXiv preprint
  arXiv:1305.0546}, 2013.

\bibitem{lions1986hopf}
P.~L. Lions and J.-C. Rochet, ``Hopf formula and multitime {Hamilton}-{Jacobi}
  equations,'' \emph{Proceedings of the American Mathematical Society},
  vol.~96, no.~1, pp. 79--84, 1986.

\bibitem{darbon2015convex}
J.~Darbon, ``On convex finite-dimensional variational methods in imaging
  sciences and {Hamilton}-{Jacobi} equations,'' \emph{SIAM Journal on Imaging
  Sciences}, vol.~8, no.~4, pp. 2268--2293, 2015.

\bibitem{ross2015primer}
I.~M. Ross, \emph{A primer on Pontryagin's principle in optimal control}.\hskip
  1em plus 0.5em minus 0.4em\relax Collegiate Publishers, 2015.

\bibitem{mitchell2005toolbox}
I.~M. Mitchell and J.~A. Templeton, ``A toolbox of {Hamilton-Jacobi} solvers
  for analysis of nondeterministic continuous and hybrid systems,'' in
  \emph{HSCC}, vol.~5.\hskip 1em plus 0.5em minus 0.4em\relax Springer, 2005,
  pp. 480--494.

\bibitem{aubin2012stochastic}
J.-P. Aubin, L.~Chen, O.~Dordan, A.~Faleh, G.~Lezan, and F.~Planchet,
  ``Stochastic and tychastic approaches to guaranteed {ALM} problem,''
  \emph{Bulletin Fran{\c{c}}ais d'Actuariat}, vol.~12, no.~23, pp. 59--95,
  2012.

\bibitem{julier2002scaled}
S.~J. Julier, ``The scaled unscented transformation,'' in \emph{American
  Control Conference, 2002. Proceedings of the 2002}, vol.~6.\hskip 1em plus
  0.5em minus 0.4em\relax IEEE, 2002, pp. 4555--4559.

\bibitem{julier2004unscented}
S.~J. Julier and J.~K. Uhlmann, ``Unscented filtering and nonlinear
  estimation,'' \emph{Proceedings of the IEEE}, vol.~92, no.~3, pp. 401--422,
  2004.

\bibitem{thrun2005probabilistic}
S.~Thrun, W.~Burgard, and D.~Fox, \emph{Probabilistic Robotics}.\hskip 1em plus
  0.5em minus 0.4em\relax MIT press, 2005.

\bibitem{ross2015unscented}
I.~M. Ross, R.~J. Proulx, and M.~Karpenko, ``Unscented guidance,'' in
  \emph{American Control Conference (ACC), 2015}.\hskip 1em plus 0.5em minus
  0.4em\relax IEEE, 2015, pp. 5605--5610.

\bibitem{ross2016path}
I.~M. Ross, M.~Karpenko, and R.~J. Proulx, ``Path constraints in tychastic and
  unscented optimal control: Theory, application and experimental results,'' in
  \emph{American Control Conference (ACC), 2016}.\hskip 1em plus 0.5em minus
  0.4em\relax IEEE, 2016, pp. 2918--2923.

\end{thebibliography}

\end{document}